\renewcommand{\v}[1]{\ensuremath{\mathbf{#1}}} 
\newcommand{\gv}[1]{\ensuremath{\mbox{\boldmath$ #1 $}}} 
\newcommand{\uv}[1]{\ensuremath{\mathbf{\hat{#1}}}} 
\newcommand{\abs}[1]{\left| #1 \right|} 
 
\let\baraccent=\= 
\renewcommand{\=}[1]{\stackrel{#1}{=}} 
\DeclareMathAlphabet{\mathsfsl}{OT1}{cmss}{m}{sl}

\documentclass[%
 reprint,
 amsmath,amssymb,
 aps,
]{revtex4-1}

\usepackage{xcolor}
\usepackage{float}
\usepackage{graphicx}
\usepackage{dcolumn}
\usepackage{bm}

\begin{document}

\preprint{APS/membrane-static}

\title{Crystalline membrane morphology beyond polyhedra}

\author{Hang Yuan}
\affiliation{%
Department of Materials Science and Engineering, Northwestern University, Evanston, IL 60208
}%

\author{Monica Olvera de la Cruz}
 \email{m-olvera@northwestern.edu}
\affiliation{
Department of Materials Science and Engineering, Northwestern University, Evanston, IL 60208
}%
\affiliation{
Department of Chemistry, Northwestern University, Evanston, IL 60208
}%
\affiliation{
Department of Physics and Astronomy, Northwestern University, Evanston, IL 60208
}%

\date{\today}

\begin{abstract}
Elastic crystalline membranes exhibit a buckling transition from sphere to polyhedron. However, their morphologies are restricted to convex polyhedra and are difficult to externally control. Here, we study morphological changes of closed crystalline membrane of super-paramagnetic particles. The competition of magnetic dipole-dipole interactions with the elasticity of this magnetoelastic membrane leads to concave morphologies. Interestingly, as the magnetic field strength increases, the symmetry of the buckled membrane decreases from 5-fold to 3-fold, to 2-fold and, finally, to 1-fold rotational symmetry. This gives the ability to switch the membrane morphology between convex and concave shapes with specific symmetry and provides promising applications for membrane shape control in the design of actuatable micro-containers for targeted delivery systems.
\end{abstract}

\maketitle

\begin{figure}[htb]
 \includegraphics[width=0.45\textwidth]{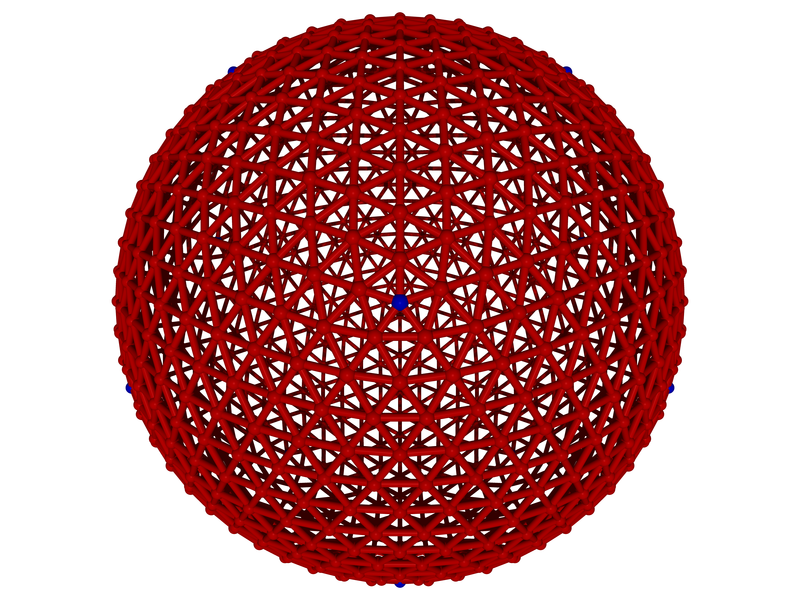}
 \caption{Mesh configuration of the spherical shell according to Caspar and Klug construction, which is characterized by two integers $h$ and $k$\cite{Caspar01011962}. Above figure shows the example of $(6,6)$ strucutre and it has 1082 vertices, 3240 edges and 2160 faces. Blue vertices correspond to the locations of 5-fold disclinations and there are 12 disclinations in total which are located on vertices of an inscribed icosahedron.}
 \label{mesh}
\end{figure}
\begin{figure}[ht]
 \includegraphics[width=0.5\textwidth]{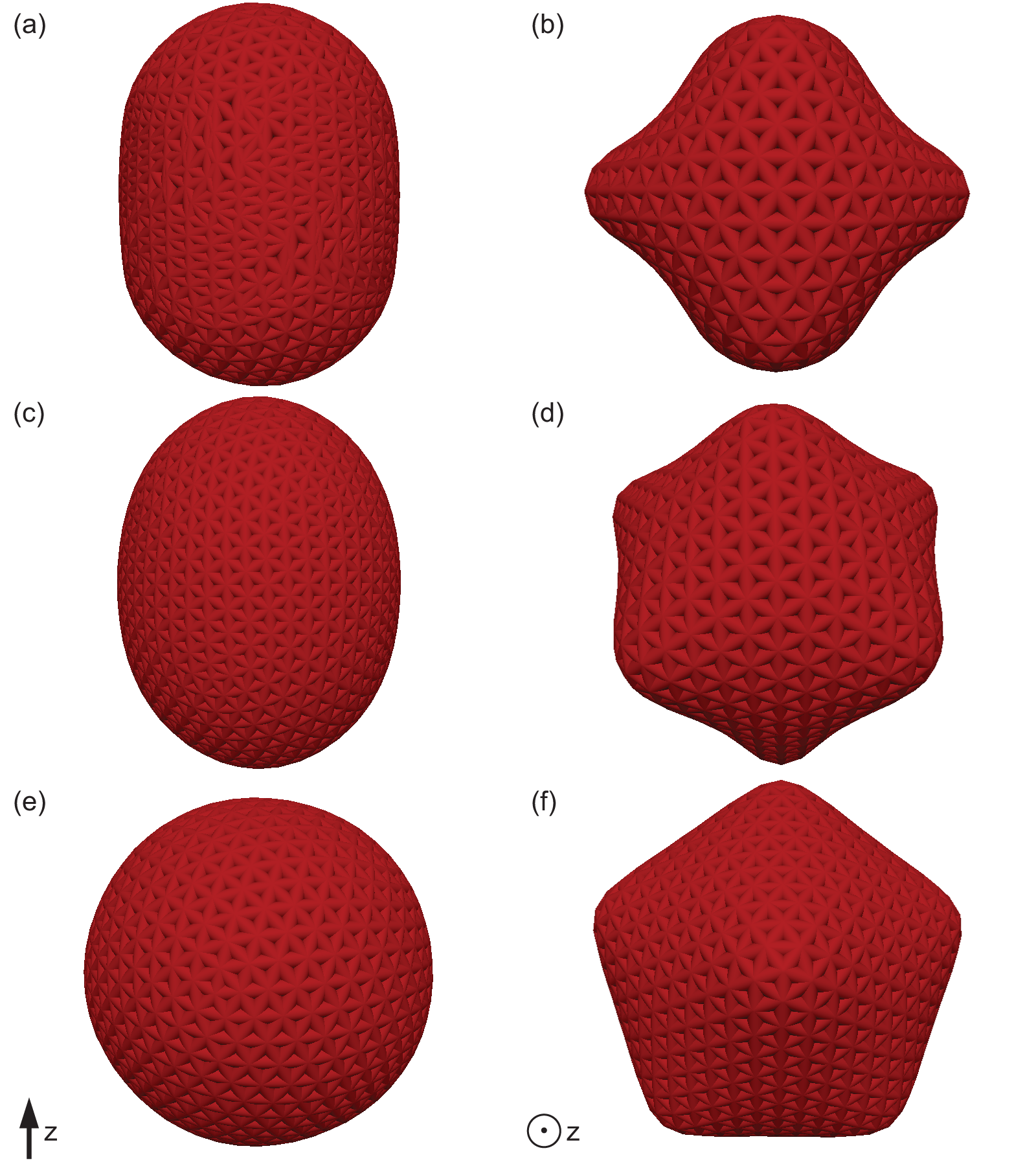}
 \caption{A collection of representative minimum energy morphologies of closed magnetoelastic membranes with different parameters pair $(\gamma,\Gamma)$: F$\mathrm{\ddot{o}}$ppl-von K\'arm\'an parameter $\gamma$ and magnetoelastic parameter $\Gamma$; $\gamma$ increases from left to right and $\Gamma$ increases from bottom to top. (a) cylindrical shape (100,50); (b) star shape with four ridges (1000,200); (c) ellipsoid shape (100,25); (d) star shape with six ridges (1000,100); (e) spherical shape (100,0); (f) icosahedral shape (1000,0). Note that first column is shown from y-direction and second column is shown from z-direction to give better illustration of morphologies. Arrows indicate the direction of the external magnetic field.}
 \label{morphologies}
\end{figure}

Polyhedra are of great interest of scientists, mathematicians and engineers. They emerge spontaneously in many fields of science. For example, single crystals take various polyhedra shapes, fullerenes adopt beautiful truncated icosahedron shapes\cite{Kroto1985}, and bacterial micro-compartments are observed in multiple regular and irregular polyhedral shapes\cite{Fan2010}. \\
\indent Self-assembled crystalline membranes, like the shells of viruses, generally possess icosahedral symmetry\cite{Zandi2004}. These icosahedral membrane shapes have been explained by homogeneous elasticity theory\cite{PhysRevA.38.1005,PhysRevE.68.051910}. Furthermore, membranes with heterogeneous elasticity have been demonstrated to form various regular and irregular polyhedral shapes\cite{Vernizzi4292}. Such polyhedral morphologies are formed by the competition between stretching energy and bending energy. Although it is possible to engineer membrane morphologies by arranging defects in closed membrane topologies\cite{PhysRevLett.111.177801}, these morphologies cannot go beyond polyhedra.\\
\indent Here, we explore the possibility to create new closed shell morphologies, other than polyhedra, in a controllable manner. For this purpose, we consider elastic membranes of super-paramagnetic particles because of the exceptional penetration of magnetic fields and bio-compatibility, which provide opportunities for biotechnology applications. Magnetoelastic materials form rich morphologies\cite{Kim2018,LumE6007} and can accomplish multimodal locomotion\cite{Hu2018} as well as deformations that generate forces between surfaces\cite{Brisbois2019} when directed by magnetic fields. The versatility of magnetoelastic filaments, which consist of super-paramagnetic particles connected by elastic linkers, has also been demonstrated experimentally\cite{Dreyfus2005,Wang2011,Wei2016} and numerically\cite{PhysRevE.95.052606,Vazquez-Montejo2017}.\\
\indent Comparing with magnetoelastic filaments and open membranes, closed magnetoelastic membranes, which have additional topological constraints, are found here to generate specific symmetries due to the interplay between nonlinear elasticity and magnetic dipole-dipole interactions. By using molecular dynamics simulations, we find the minimum energy configurations of magnetoelastic membranes, which can be directly controlled by external magnetic fields.\\
\indent As dictated by Euler's polyhedron formula, we start by triangulating a spherical shell with twelve isolated 5-fold disclinations. The disclinations are positioned on the vertices of an inscribed icosahedron (Fig. \ref{mesh}) to minimize the interactions between them\cite{PhysRevB.62.8738}, as proposed by Caspar and Klug\cite{Caspar01011962}.\\
The elastic component of the Hamiltonian of a magnetoelastic membrane, following the discretization scheme of Nelson et al\cite{PhysRevA.38.1005}, is written as
\begin{equation}
H_e=\sum_{e\in \v E}\frac{1}{2}k\left(\abs{\v r_1^e - \v r_2^e} - l_0\right)^2 + \sum_{e\in \v E} \frac{1}{2}\tilde \kappa \abs{\v n_1^e - \v n_2^e}^2
\end{equation}
where $k$ is microscopic stretching constant and $\tilde \kappa$ is microscopic bending rigidity. The sum is over all $e$ elements of $\v E$, which is the set of all edges; $\v r_1^e$ and $\v r_2^e$ are two vertices of the edge $e$; and $\v n_1^e$ and $\v n_2^e$ are normal vectors of the two adjacent triangles of the edge $e$; and $l_0$ is the equilibrium length. Note that the corresponding continuum limit of the above discretized Hamiltonian is mesh dependent\cite{Gompper1996}. With the above described triangulation of a spherical shell, it has been shown that in the continuum limit\cite{PhysRevA.38.1005,Nelson2004} Young's modulus $Y=\frac{2k}{\sqrt 3}$, Poisson's ratio $\nu =\frac{1}{3}$ and bending rigidity $\kappa=\frac{\tilde \kappa}{\sqrt 3}$.\\
\indent Incompressible membranes ($\nu=1/3$) of radius $R$ can be described by two parameters $Y$ and $\kappa$. Then, a single dimensionless parameter, $\gamma=\frac{YR^2}{\kappa}$, called the F$\mathrm{\ddot{o}}$ppl-von K\'arm\'an parameter\cite{Libai1998}, completely determines the buckling transition of the system. Nelson et al\cite{PhysRevE.68.051910} have shown that homogeneous elastic membranes undergo a spontaneous buckling transition from sphere to icosahedron when $\gamma > \gamma^\ast=154$, where $154$ is the value of $\gamma^\ast$ for a flat disk.\\
\indent In our study, we place a small super-paramagnetic particle at each vertex. An external magnetic field induces a magnetic dipole on each vertex. Therefore, an additional term from magnetic dipole-dipole interactions is added into the Hamiltonian of the system:
\begin{equation}
H_m=-\frac{\mu_0}{4\pi}\sum_{\v r_i,\v r_j\in \v V}\frac{1}{\abs{\v r_{ij}}^3}\left[3\left(\gv \mu_i \cdot \uv r_{ij}\right)\left(\gv \mu_j\cdot \uv r_{ij}\right)-\gv \mu_i\cdot \gv \mu_j\right]
\end{equation}
where $\mu_0$ is the magnetic permeability in vacuum, $\gv \mu_i$ is the magnetic dipole moment at vertex $i$, $\v V$ is the set of all vertices, $\v r_i$ is the position vector of vertex $i$, $\v r_{ij}=\v r_j - \v r_i$ and $\uv r_{ij}=\v r_{ij}/\abs{\v r_{ij}}$ and the sum is over $i\neq j$.\\
\indent The magnetic dipole-dipole interaction is long range and anisotropic. A useful simplified form which considers only nearest neighbor interactions in the inextensible limit\footnote{For details of derivation, please refer to SI.} is derived and yields:
\begin{equation}
H_m\approx \left(\sum_{\v r_i \in \v V^{hex}}6 + \sum_{\v r_i \in \v V^{pen}}5\right)\left({n_z^{i}}^2 - \frac{1}{3}\right)\tilde M
\end{equation}
where $\tilde M=\frac{1}{4}\frac{\mu_0}{4\pi}\frac{(3\mu)^2}{l_0^3}\frac{2}{3}$, $\mu$ is the induced magnetic dipole moment which assumes only one type of super-paramagnetic particles, $\v V^{hex}$ is the set of vertices with 6 neighbors, $\v V^{pen}$ is the set of vertices with 5 neighbors and $n_z^{i}$ is the z component of normal vector at vertex $i$.\\
$\tilde M$ gives the characteristic energy scale for each nearest neighbor pair of magnetic dipole-dipole interactions in the discretization limit. Similar to the case of elastic membranes, a magnetic modulus can be defined in the continuum limit as $M=8\sqrt 3 \frac{\tilde M}{l_0^2}$, and a dimensionless parameter, $\Gamma = \frac{MR^2}{\kappa}$, called magnetoelastic parameter\cite{PhysRevE.98.032603}, can be similarly defined. The magnetoelastic parameter $\Gamma$ characterizes the relative strength between magnetic energy and bending energy(see SI).\\
\indent Therefore, the magnetoelastic membrane has one additional energy competition from magnetic dipole-dipole interactions, which is tunable via an external magnetic field. The total magnetoelastic energy of the membrane $H_{em}$ is the sum of elastic and magnetic energies, which divided by $\kappa$ gives the dimensionless form:
\begin{equation}
\begin{aligned}
\tilde{H}_{em}\left[\{\v r_i\};\gamma,\Gamma\right]&=\frac{H_{em}\left[\{\v r_i\}\right]}{\kappa}\\
&=\tilde{H}_e\left[{\{\v r_i\}};\gamma\right]+\tilde{H}_m\left[{\{\v r_i\}};\Gamma\right]
\end{aligned}
\end{equation}
where tilde indicates dimensionless quantites.\\
\indent Besides magnetic and elastic contributions, a volume constraint is also imposed on the membranes to account for internal pressure. This internal pressure is necessary when the membrane is not penetrable, which is modeled as
\begin{equation}
    H_v=\Lambda\left(\sum_k \Omega_k - V_{ref}\right)^2
\end{equation}
where $\Omega_k$ is the signed volume of the tetrahedron extended by k-th triangle on the membrane, $V_{ref}$ is the reference volume of the membrane and $\Lambda$ is the Lagrange multiplier which serves role of pressure. $V_{ref}$ is set as volume of the icosahedron after buckling and $\Lambda$ is set to a large enough value such that the membrane has additional rigidity from the volume constraint. The volume constraint is used to better capture effect from the environment surrounding the magnetoelastic membrane and eliminate possible crumpled states\cite{RevModPhys.79.643}. Corresponding cases without the volume constraint are also explored, and their morphologies generally do not differ significantly from the cases with the volume constraint. Some crumpled states and collapsed states are observed in high field strength limit for the cases without the volume constraint(see SI for more discussions).\\
\indent In the simulation, a shifted Lennard-Jones potential is also included for each pair of vertices to account for the exclude volume effect. Each vertex is assigned a point magnetic dipole moment. Stretching and bending are treated with a harmonic bond interaction and a harmonic dihedral interaction, respectively. Magnetic dipole-dipole interactions are calculated without a cutoff. The connectivity of the membrane is preserved during simulations. The external magnetic field is static along z direction. We assume super-paramagnetic particles respond to an external magnetic field instantaneously and ignore rotational degrees of freedom of each vertex because super-paramagnetic particles do not have spontaneous magnetization, which decouples magnetics and elasticity. Kinetic energy is also assigned to each vertex to give a fictitious temperature of the system. The simulations start at high temperature and are gradually annealed to find the minimum energy configuration of the system. This annealing process is repeated several times to ensure that the system is not trapped in local minima.\\
\indent A collection of possible morphologies of magnetoelastic membranes obtained by systematically varying the two dimensionless parameters, the F$\mathrm{\ddot{o}}$ppl-von K\'arm\'an parameter $\gamma$ and the magnetoelastic parameter $\Gamma$, are shown in Fig. \ref{morphologies}.
Without magnetic dipole-dipole interactions ($\Gamma$=0), when $\gamma<\gamma^\ast$, the homogeneous elastic membrane tends to stay spherical (Fig. \ref{morphologies}e) and when $\gamma>\gamma^\ast$ it buckles into an icosahedron (Fig. \ref{morphologies}f) as expected in the conventional homogeneous elastic membranes\cite{PhysRevE.68.051910}.\\
\indent At moderate strengths of the magnetic dipole-dipole interaction, as shown in second row of Fig. \ref{morphologies}, the structures deform since the magnetic dipoles prefer to line up and stay closer to each other to minimize the magnetic energy. When the membrane is relatively soft ($\gamma < \gamma^\ast$), the membrane tends to elongate along the direction of the external magnetic field. However, this is opposed by elastic interactions since elasticity prefers the membrane to stay spherical, resulting in an ellipsoid like membrane morphology as shown in Fig. \ref{morphologies}c.\\
When the membrane is relatively stiff ($\gamma > \gamma^\ast$), the membrane undergoes an elastically driven buckling transition. The interplay between nonlinear elasticity and magnetic dipole-dipole interactions distorts the icosahedron. The flat regions of the icosahedron bend inward to reduce the distance between magnetic dipoles and disclinations pair up, resulting in a star-like morphology with six ridges as shown in Fig. \ref{morphologies}d. Unlike the conventional convex polyhedral morphologies of the purely elastic membranes, the magnetoelastic membranes develop concave regions.\\
\begin{figure*}[htb]
 \includegraphics[width=\textwidth]{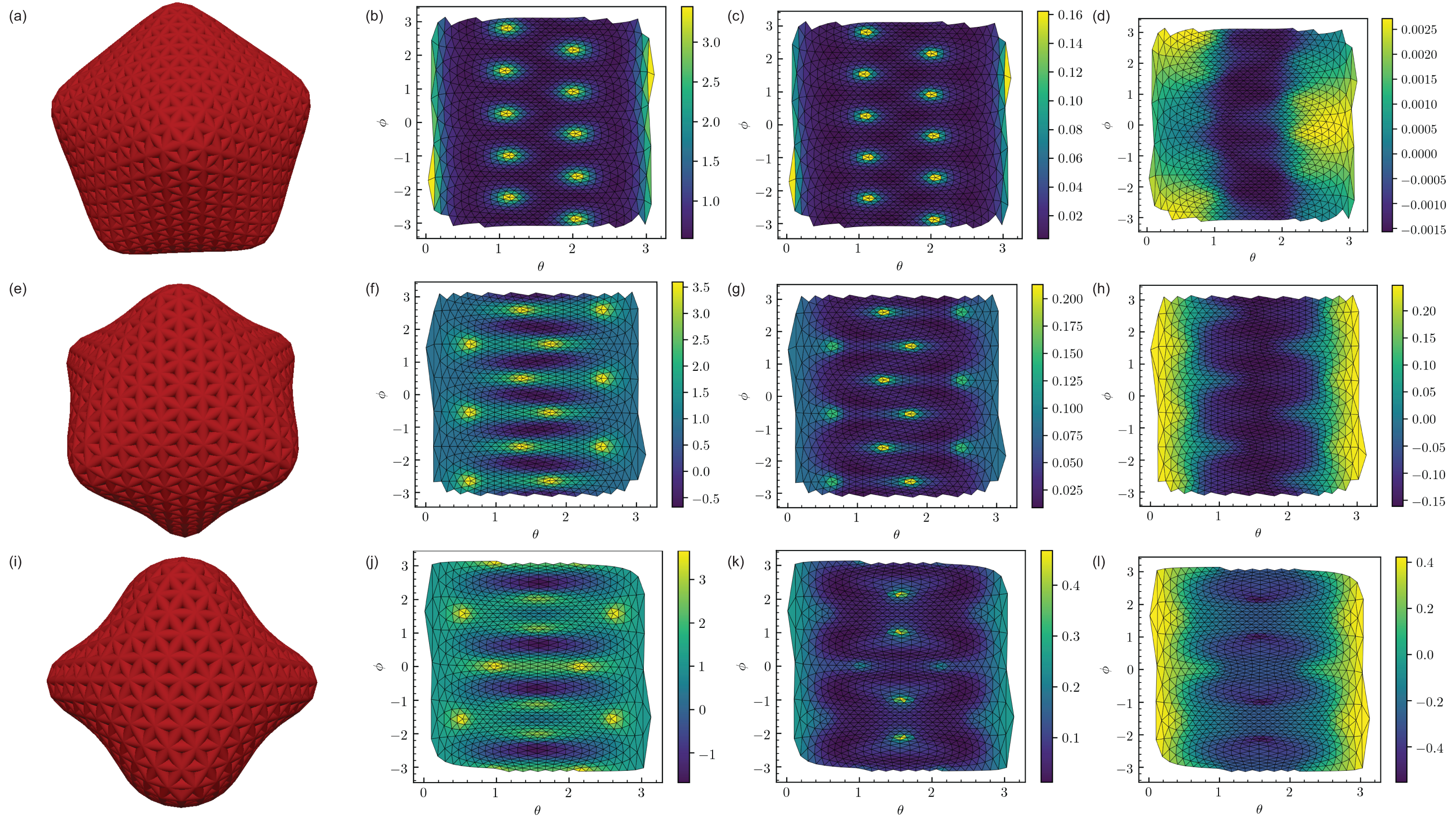}
 \caption{membrane morphologies(first column), mean curvature distribution(second column), elastic energy distribution(third column) and magnetic energy distribution(fourth column) showing symmetry of magnetoelastic membranes. Membrane morphologies are all shown in z-direction. All other plots are shown in spherical coordinates. Horizontal axis is polar angle $\theta \in [0,\pi]$ and vertical axis is azimuthal angle $\phi \in (-\pi, \pi]$. Mean curvatures are chose to be signed values where positive values indicate convex regions and negative values indicate concave regions. Elastic energy is sum of stretching energy(bond interaction) and bending energy(dihedral interaction). Magnetic energy is sum of magnetic dipole-dipole interactions and Lennard-Jones energy is negligible in all three cases. Parameters pairs $(\gamma, \Gamma)$ of membranes in each row are: (1000,1), (1000,100), (1000,200) from top to bottom.}
 \label{curvatures}
\end{figure*}
Then, consider the case of strong magnetic dipole-dipole interactions, as shown in first row of Fig. \ref{morphologies}. The elastic energy only becomes comparable with the magnetic energy until the membrane is highly deformed. In this regime, the competition between magnetic energy and elastic energy results in another new family of morphologies.\\
\indent When the membrane is easily deformed ($\gamma < \gamma^\ast$), magnetic dipole-dipole interactions tend to elongate the membrane furthermore along the direction of external magnetic field in this high field strength regime. However, the elastic energy can no longer hold the membrane in a spherical or ellipsoidal shape. The membrane forms a cylindrical shape, as shown in Fig. \ref{morphologies}a, to minimize the magnetic energy. Although the bending energy is high along edges of two end caps of the cylinder, the total energy decreases by lining up vertices on the side surface of the cylinder.\\
\indent When the membrane is relatively rigid ($\gamma > \gamma^\ast$), the elastic energy tries to preserve the total surface area of the membrane since stretching is much more expensive than bending in this case. Meanwhile, the magnetic dipole-dipole interaction tries to reduce the total volume of the membrane to minimize the magnetic energy. This competition, combined with the nonlinearity introduced by the twelve disclinations, results in a star-like morphology with four ridges as shown in Fig. \ref{morphologies}b. Note that the membrane in this case is highly bent inward, which reduces its total volume significantly and opens some possible applications as discussed later.\\
\indent Among all these mentioned morphologies of the magnetoelastic membrane, the $\gamma\sim 1000$ cases are particularly interesting because this regime corresponds to a typical F$\mathrm{\ddot{o}}$ppl-von K\'arm\'an parameter of viral shells\cite{PhysRevE.68.051910}. In this regime, where both nonlinear elasticity and magnetic dipole-dipole interactions are significant, we find that the magnetoelastic membrane tends to choose configurations that decrease symmetry with increasing external magnetic field strength. This point is illustrated by plotting the mean curvature and energy distribution of the membrane in spherical coordinates as shown in Fig. \ref{curvatures}.\\
\indent In the weak field strength limit, the membrane forms an icosahedron (Fig. \ref{curvatures}a) with five-fold rotational symmetry around the z-axis as shown in Fig. \ref{curvatures}b and \ref{curvatures}c. In this limit, the elastic energy dominates and the magnetic energy is negligible (Fig. \ref{curvatures}d). With a moderate external magnetic field strength, the membrane morphology has six ridges(Fig. \ref{curvatures}e). However, the 12 isolated disclinations prefer to pair up and form ridges connecting each pair of disclinations\cite{Lobkovsky1995}. These disclinations pairs are arranged alternatively to maximize the mutual distance in order to reduce interactions between disclinations\cite{PhysRevB.62.8738} and ridges\cite{PhysRevE.55.1577}, as shown in Fig. \ref{curvatures}f and \ref{curvatures}g. Because of this alternative arrangement, the membrane with six ridges has only three-fold rotational symmetry around the z-axis, which is also reflected by magnetic energy distribution (Fig. \ref{curvatures}h). If the field strength is further increased, the membrane starts to form morphologies with four ridges (Fig. \ref{curvatures}i). In this regime, two pairs of disclinations break and there is a single disclination near each of the four concave regions as show in Fig. \ref{curvatures}j and \ref{curvatures}k. Therefore, the symmetry of the membrane reduces to two-fold rotational symmetry (Fig. \ref{curvatures}j, \ref{curvatures}k and \ref{curvatures}l) around the z-axis. In the extremely high field strengths regime, the magnetic energy completely dominates and the membrane collapses and takes one-fold rotational symmetry(the collapsed state is not shown in Fig. \ref{curvatures}).\\ 
\indent A natural question is to ask why the four-fold rotational symmetry is missing among all these above mentioned morphologies. A qualitatively argument suggests that if the system was to form a morphology with four-fold rotational symmetry, the 12 disclinations must be divided into groups of 3 disclinations, which means pairs of disclinations must break up. A direct visualization of the four-fold rotational symmetry case is to image the two-fold rotational symmetry case without alternative arrangement of the disclinations. Since weak field strengths cannot break up disclinations pairs and strong field strengths prefer alternative arrangement of disclinations to lower the system energy, the four-fold rotational symmetry is missed in these morphologies.\\
\indent When the volume constraint is removed, we find similar morphologies to those discussed above except in high magnetic field strengths, where they take crumpled or collapsed morphologies with one-fold rotational symmetry(see Fig.2 in SI).\\
\indent In summary, crystalline magnetoelastic membranes exhibit concave morphologies beyond the conventional polyhedral shapes found in elastic membranes. Magnetic dipole-dipole interactions give an additional control parameter which is characterized by the magnetoelastic parameter $\Gamma$. Combining with the F$\mathrm{\ddot{o}}$ppl-von K\'arm\'an parameter $\gamma$ in the elastic membranes, these two dimensionless parameters provide guidelines for analyzing properties of crystalline magnetoelastic membranes. Importantly, since $\gamma$ is hard to change once a membrane is assembled, the magnetoelastic parameter, which can be easily manipulated by an external magnetic field, provides a way to tune the membrane morphology between convex shapes and concave shapes with specific symmetry.\\
\indent Exciting applications, including reversible membrane shape control, design of micro-container, and targeted drug delivery, are expected for the closed crystalline magnetoelastic membranes. For example, since the volume to surface ratio of magnetoelastic membranes can be highly reduced by imposing an external magnetic field, the concentration inside can be much higher than that in the outside environment. This morphological change induced by the external magnetic field can facilitate release of cargoes. Therefore, the magnetoelastic membrane can be used as a container to carry and protect volatile or toxic molecules and release them in targeted region labelled by the external magnetic fields.
\begin{acknowledgments}
This work was supported as part of the Center for Bio-Inspired Energy Science, an Energy Frontier Research Center funded by the US Department of Energy, Office of Science, Basic Energy Sciences under Award DE-SC0000989.
\end{acknowledgments}



\bibliographystyle{unsrt}
\bibliography{Literatures}

\end{document}